\documentclass[twocolumn,10pt,prl,nofootinbib]{revtex4}

\def\mysection#1{{\bf #1.} }

\usepackage[dvips]{graphicx}
\usepackage{epsfig,amsmath,amssymb,verbatim,mathrsfs,array,layout,textcomp,amssymb,latexsym, slashed,graphicx,bbm}

\usepackage{graphicx}
\usepackage{amsmath}
\usepackage{amssymb}
\usepackage{appendix}
\usepackage{verbatim,mathrsfs,array,layout,textcomp,latexsym, slashed}

\usepackage{amsmath, amssymb, graphicx, rotating}
\usepackage{hyperref}
\usepackage[usenames,dvipsnames]{color}

\usepackage{bm}

\newcommand{\bea}{\begin{eqnarray}}
\newcommand{\eea}{\end{eqnarray}}

\newcommand{\no}{\nonumber}
\newcommand{\be}{\begin{equation}}
\newcommand{\ee}{\end{equation}}

\newcommand{\MeV}{\text{ MeV}}
\newcommand{\eV}{\text{ eV}}
\newcommand{\keV}{\text{ keV}}

\def\lsim{\mathrel{\rlap{\lower4pt\hbox{\hskip1pt$\sim$}}
     \raise1pt\hbox{$<$}}}         
\def\gsim{\mathrel{\rlap{\lower4pt\hbox{\hskip1pt$\sim$}}
     \raise1pt\hbox{$>$}}}         

\begin{document}

\font\mini=cmr10 at 0.8pt

\title{
Sub-GeV Dark Matter Detection with Electron Recoils in Carbon Nanotubes
}

\author{G. Cavoto}\email{gianluca.cavoto@roma1.infn.it}
\author{F. Luchetta}
\author{A.D. Polosa}
\affiliation{Sapienza University of Rome, Piazzale Aldo Moro 2, I-00185 Rome, Italy}

\begin{abstract}
Directional detection of Dark Matter particles (DM) in the MeV mass range could be accomplished by studying  electron recoils in  large arrays of parallel carbon nanotubes.  In a scattering process with a lattice electron, a DM particle might transfer sufficient energy to eject it from the nanotube surface. An external electric field is added to drive the electron from the open ends of the array to the detection region.  The anisotropic response of this detection scheme, as a function of the orientation of the target with respect to the DM wind, is calculated, and it is concluded that no direct measurement of  the electron ejection angle is needed to explore significant regions of the light DM exclusion plot.  A compact sensor, in which the cathode element is substituted with a dense array of parallel carbon nanotubes, could serve as the basic detection unit. 
\end{abstract}

\maketitle

\mysection{Introduction}\label{sec:intro}
Two-dimensional targets for directional dark matter searches have been recently studied in~\cite{Capparelli:2014lua} and~\cite{Hochberg:2016ntt}. Array of carbon nanotubes considered in~\cite{Capparelli:2014lua} could work as highly transmitting channels for carbon ions recoiled by DM particles  with masses $M_\chi> 1$~GeV.  If nanotube axes are aligned in the direction of the Cygnus constellation, along which the largest fraction of the DM  velocity vectors are oriented, a significantly higher number of carbon ions  is expected to be channeled with respect to the case in which the axes are rotated by 180$^\circ$.  Interstices among carbon nanotubes (CNT) are also  found to cooperate to enlarge the effective channeling angle of the array: the maximal recoil angle a carbon ion can have to be  channeled by the array and eventually be detected is computed in~\cite{Cavoto:2016lqo}.

If   instead electron recoils  are considered, graphene sheets are potentially very good directional targets for DM in the MeV mass range~\cite{Hochberg:2016ntt}.  Most of the ideas on sub-GeV DM and on the  possibilities  for exploring and revealing it are summarized in the report~\cite{Alexander:2016aln};  more specifically see Refs.~\cite{Essig:2011nj}--\cite{Hochberg:2015fth}.

The electrons or ions recoiling against the hitting DM particles on two-dimensional layers are emitted by the material, with reduced internal rescatterings, differently from crystalline  or gaseous targets.

The energy price to pay to extract valence band, $\pi$-orbital electrons from graphene is in the order of few eVs (the work function being $\phi_{\rm wf}\approx 4.3\eV$ as opposed to a minimum of $20\eV$ to eject a  carbon atom\footnote{In addition to this, more energy is needed to eject an ionized carbon nucleus, necessary condition to be channeled in the nanotube array. }). Due to the nature of the scattering with electrons in the graphene structure,  {\it electron recoils tend to follow the same direction of the incident DM}.  Graphene layers, oriented perpendicularly to the DM wind, tend to emit electrons in the same direction, which should be immediately collected/detected. 
A measurement of the recoil angle would provide clear directional information enhancing the capabilities of  background rejection.

In this note we follow the suggestion by Hochberg {\it et al.}~\cite{Hochberg:2016ntt} of using electron recoils from both $\pi$ and $sp^2-$orbitals in graphene,  but again we resort to the wrapped configuration provided by carbon nanotubes ({\it single-wall} carbon nanotubes are essentially graphene sheets wrapped on a cylindrical surface). This allows to reach a higher density of target material, {\it i.e.} smaller detectors, which in turn could be more easily handled and oriented in the DM wind direction. With carbon nanotubes we find  the same directional behavior of electron recoils it is found in~\cite{Hochberg:2016ntt} for graphene layers.  

\mysection{The target scheme}\label{sec:sheme0}
When electron recoils are considered, the nanotube walls cannot work as  reflecting surfaces, as they were considered in~\cite{Capparelli:2014lua,Cavoto:2016lqo}, capable of channeling ions  having  transverse energies lower than the reflecting potential barriers at the 
boundaries (in the order of few hundreds of eVs).  On the contrary, electrons injected in the body of nanotubes (or in the interstices) by  DM-electron scatterings,  might go through the walls, with definite  transmission coefficients, and undergo multi-scattering events, crossing several nanotubes, before exiting from the array. 
As a benchmark for our analysis, we have used the results of some experimental studies on the determination of transmission coefficients of electrons through graphene planes~\cite{Hassink, Longchamp, Yan, Mutus, Kim}.  More information from experiments of this kind would be extremely useful for determining directly also reflection and absorption coefficients. 

\begin{figure}
\centering
\includegraphics[width=7.5truecm]{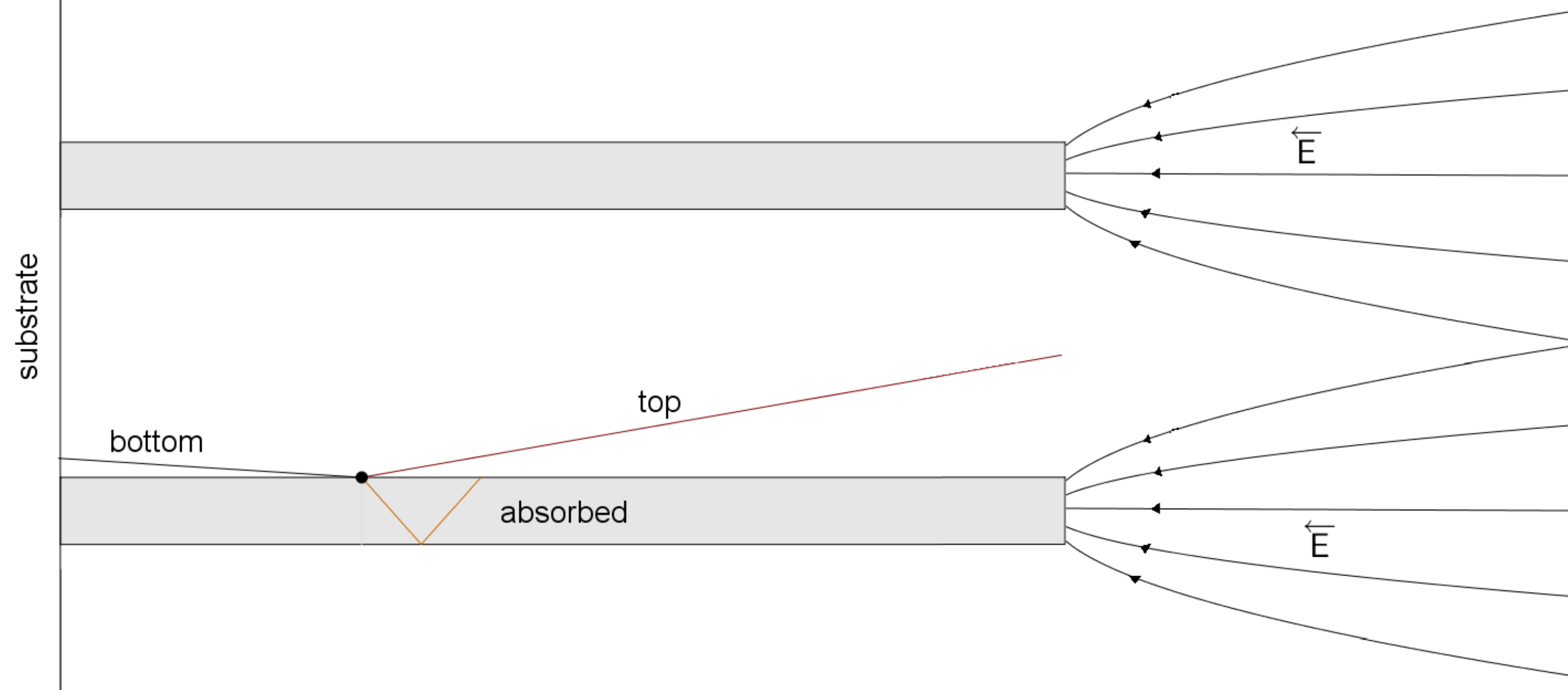}
\caption{\small Scheme of the basic unit of the carbon nanotube array. The grey sectors represent the sections of two close-by nanotubes. 
 The detection apparatus is located on the side of the open ends of the array (on the right of the scheme). Electrons can also be transmitted through the nanotubes walls or reflected/absorbed. }
\label{fig:scheme}
\end{figure}

The  addition of  an electric field $\bm E$, coaxial with  nanotube parallel axes works to drive the ejected electrons to the detection region in the direction opposite to the substrate, where the nanotubes have been deposited on --- see Fig.~\ref{fig:scheme}. Following~\cite{Capparelli:2014lua,Cavoto:2016lqo}, we  consider to align the nanotube axes in the DM wind direction in order to get most of the recoils in that direction. The alignment can be kept fixed by a continuous mechanical tracking system. 

We assume that the carbon nanotube array is engineered as a forest of  metallic nanotubes, on a conducting plate.  An opposite electrode makes an electric field  $E$  directed to the former with field lines concentrated as on sharp edges, at the nanotube ends. If $R$ is the average distance between the axes of two nanotubes in a square array and $r<R$ is the nanotube radius, the electric field intensity  will increase at the extremity as in
\be
E^\prime\approx\frac{1}{2}\frac{R^2}{r^2}E
\ee
With $R\simeq 50$~nm and $r\simeq 5$~nm, this might allow to reach an electric field of $E^\prime\approx 500$~kV/cm at the ends of the nanotubes with $E\approx 10$~kV/cm -- the typical electric fields used to collect ejected electrons (see discussion below).  $E^\prime$ must not be large enough to produce field emission electrons, a potentially important background. As from the Fowler-Nordheim theory of field emission from metallic carbon nanotubes (see for example~\cite{fowler}), the characteristic field emission currents are
\be
\label{field}
j(\mu A)\simeq 7.5\frac{E^\prime}{\phi^{1/2}}\exp\left(-6.83\frac{\phi^{3/2}}{E^\prime}\right) \coth\left(5.6 \frac{\phi^{1/2}}{2E^\prime r}\right )
\ee
where $E^\prime$ is expressed in V/nm, the work function is $\phi\simeq 4$~eV and the radius of the nanotube $r$ is in nm.  The expected current per nanotube at 500 kV/cm is negligibe, $j\approx \exp(-10^3)~\mu A$, even when multiplied by the whole number of nanotubes in the array.
We observe here that a further reduction of the electric field at the nanotube ends 
can be reached by  decreasing the average relative distance $R$ among the seeds on which the nanotubes are grown.


As reminded above, electron recoils are mainly forward, keeping track of the DM direction.
We call $N_+=N(\theta_w)$ the number  of electrons reaching the detection region as a function of the angle $\theta_w$ between the average direction of the DM wind and the carbon nanotube (parallel) axes, oriented along the direction from the closed bottom to the open ends.  We also define $N_-=N(180^\circ)$ and we will seek for $\theta_w$ angles giving the largest asymmetry 
\begin{equation}
A(\theta_w) = \frac{N_+ - N_-}{N_+ + N_-}
\label{asymmetry}
\end{equation}

On the basis of the results obtained in~\cite{Hochberg:2016ntt}, we expect $A$ to be maximal in correspondence of $\theta_w\simeq 90^\circ$
where the emission cross section is higher. However, the electrons which will most likely reach the driving electric field are those recoiled along the axes of nanotubes --- see the discussion below on the low energy transmission through carbon nanotubes.

According to  our simulations, an asymmetry as large as $A\sim 0.4$ can be reached. Such a value of $A$  could be measured  with $5\sigma$ experimental significance by counting a total number of about 60 events, in absence of background.

To observe the anisotropy $A$,  there is no need to measure the  ejection angle of  recoiling electrons. Only an efficient electron counting and mechanical tracking system  is needed. This might allow to consider a detection apparatus in which, for instance,   the carbon nanotubes array target is replacing  cathode of a compact device. 
We will illustrate in what follows how we reach these conclusions.  

\mysection{Trajectories and the absorption coefficient}\label{sec:sheme}
The basic unit of a carbon nanotube array is sketched in Fig.~\ref{fig:scheme}. A collision with a DM particle might generate a `top' electron, which aims in the direction towards the  open ends of  the nanotubes, where the detection apparatus is located,  or a `bottom' electron, which is instead  directed towards the substrate, where it is always absorbed. Top electrons might be reflected, transmitted or just stopped/absorbed by the nanotube walls. 

In the conditions described, electrons are in the  the 1-10~eV energy range, {\it i.e.}, they have negligible  resolution power for atoms and nuclei. These long wavelength electrons (between 4 and 12 \AA) interact with portions of the graphene (or nanotube surfaces)  producing a diffraction pattern in transmission, as discussed in~\cite{nature} and~\cite{ucla}. The largest intensity is expected in the forward direction, as can be seen from Fig.~9 in~\cite{ucla}  and Fig.~4 in~\cite{nature}. Secondary maxima are found at angles $\theta$ between the incident electron wave-vector $\bm k$ and the final one $\bm k^\prime$ given by $\sin(\theta/2)=\big(\lambda/3l\big)\,\sqrt{m_1^2+m_2^2+m_1m_2}$ in the elastic approximation $|\bm k|\sim|\bm k^\prime|$ --- here $l$ is the bound length $l\simeq0.14$~nm and $m_{1,2}$ are integers.  
\newline\indent
Almost everywhere in the range $1<E<10$~eV, the previous equation has only the solution $m_1,m_2=0$ and $\theta=0$, corresponding to 
forward transmission. In the upper part of the energy range $E\gtrsim 8$~eV, other solutions are possible but one has to choose those not involving a change of  the $z$ component since $\bm k^\prime-\bm k=\bm q$ with $\bm q$ in the graphene plane. To the best of our knowledge, there are no experiments giving precise information
about the relative weight of transmissions versus reflections from 
graphene monolayers. Specular reflections of low energy electrons are expected to occur at low energies. 
\newline\indent
Significant changes in the size of momentum, $|\bm k^\prime|\ll |\bm k|$, correspond to `absorptions'.
The condition $| (k^2 - q^2) / 2k^2| \ll 1$ ($\bm q$ is the momentum exchanged with the lattice) suggests that electrons grazing the surface of nanotubues can be more easily absorbed.
\newline\indent
Indeed in the experiments reported  in~\cite{Hassink, Longchamp, Yan, Mutus, Kim} it is found that the largest part of the electron beam impinging orthogonally to the graphene monolayer (deposited on a surface with holes) is transmitted  or reflected. 
\newline\indent
The probability of transmission $T$, reflection $R$ and absorption $C$ are introduced ($T+R+C=1$). In the following,  as suggested in~\cite{Hassink, Longchamp, Yan, Mutus, Kim} we will keep $T+R$ to be the dominant fraction
thus varying only the small $C=1-(T+R)$ value. The chosen values of $C$ are suggested in the quoted references; in particular in~\cite{Kim}, $C$ is measured to be $C\sim 10^{-4}$ for electrons with $E\lesssim 3$~eV.  

The simulation of trajectories follows a standard procedure. 
A random nanotube in the array and a random point $P$ on it are chosen. An electron is ejected from $P$ with some random direction. 
Electrons facing the substrate/open-ends are  labelled as the `bottom'/`top' ones. 
Positions and velocities can be computed at every step of the simulation (using an Euler algorithm). This allows to reconstruct the whole trajectory. At every intersection of the trajectory with any nanotube in the array, a transmission/reflection/absorption is decided with  probabilistic weights $T,R,C$. Some few trajectories terminate neither on the side where the open ends of the nanotube array are, nor on the substrate: there is a limited number of `side' electrons, similarly to what was found for `side'-ions in the simulations discussed in~\cite{Cavoto:2016lqo}. 
The calculation are repeated for an arbitrary number of initial  electrons.  

The average distance spanned by recoiled electrons in quasi-parallel directions to nanotube axes (within $0^\circ\div 10^\circ$),  is several hundreds $\mu$m,  in the range of absorption coefficients  we are considering | the length of aligned carbon nanotubes being $\approx 200~\mu$m.

\mysection{Results}
\begin{figure*}[ht!]
\centering
\includegraphics[width=9truecm]{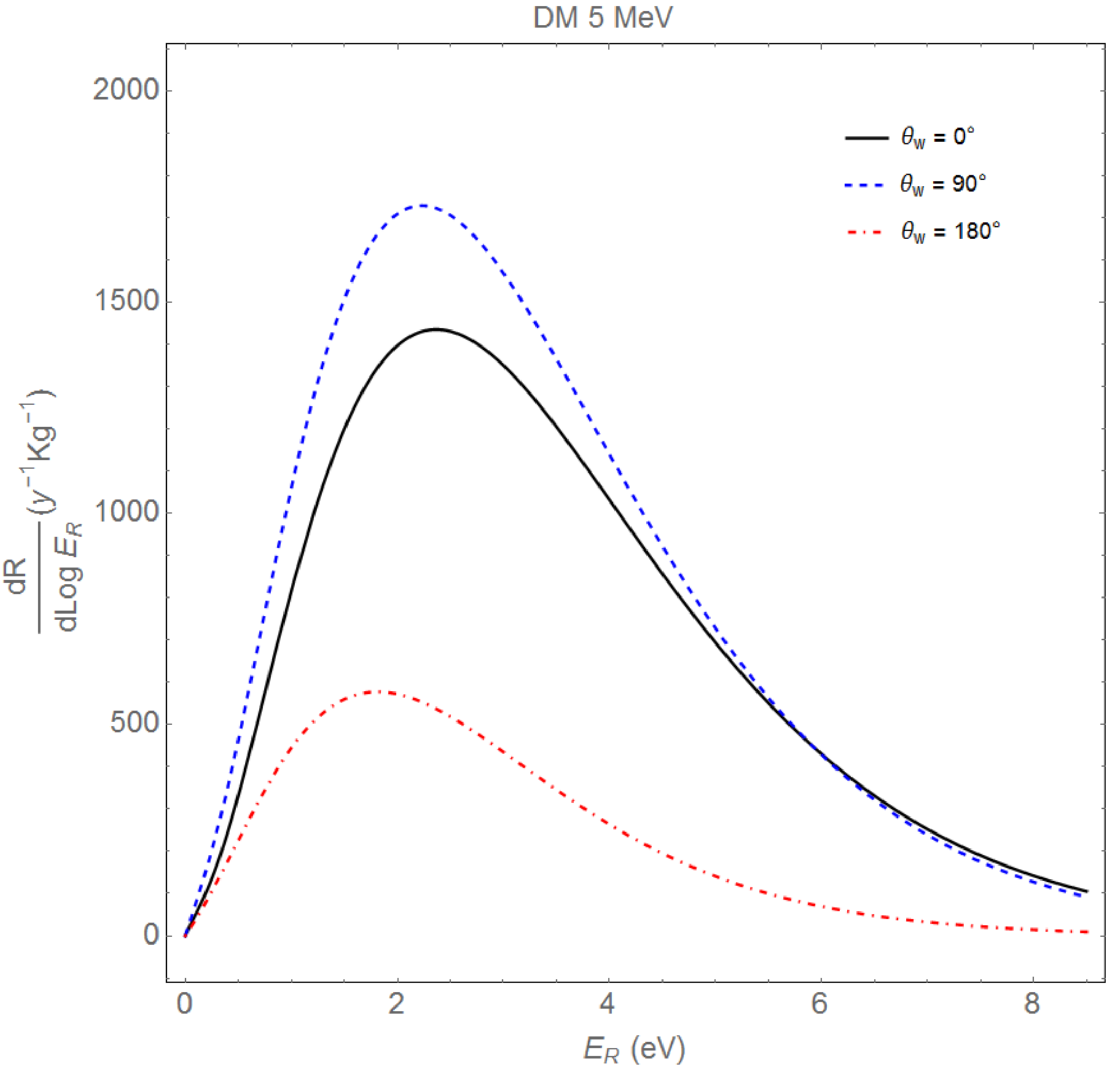}
\caption{\small Differential rates of ejected electrons per year per kg, distributed in the recoil energy $E_R$ for $M_\chi=5$~MeV.  We include both $sp^2$ and $\pi$--orbital electrons. The three curves reported  are relative to three different orientations of the DM wind main direction with respect to the carbon nanotube parallel axes.   The plot reported here is found with Eq.~\eqref{integ}  and the addition of the absorption probability at every hit with the CNT.  $E_R$ corresponds to the kinetic energy of the electrons emitted from the surface of the CNTs, having overcome the work function $\phi_{\rm wf}$. 
}\label{grafici_rate}
\end{figure*}
Following~\cite{Hochberg:2016ntt}, we consider now the collision of DM particles with graphene electrons in both  $\pi$ and $sp^2-$orbitals,  with cross section given by Eq.~\eqref{xsect}, in the Appendix. This depends on the $|\widetilde\psi(\bm q-\bm k^\prime,\bm\ell)|^2$ factor, which measures the probability for the recoiled electron to have 3-momentum $\bm k^\prime$ when the value of the exchanged 3-momentum with the graphene lattice is $\bm q$ and $\bm \ell$ is the lattice momentum in the Brillouin zones. In our description, single-wall carbon nanotubes correspond to wrapped graphene planes. Thus, first of all,  we have reproduced, with perfect agreement,  the results illustrated in~\cite{Hochberg:2016ntt} and then considered periodic boundary conditions on graphene planes to get the appropriate $\widetilde\psi$ functions.

The diameter of a nanotube is of the order of $2R=100$~\AA~and we consider electrons having recoil kinetic energies in the range\footnote{There are not many papers on the experimental determination of transmission coefficients of  graphene at these energies. Hassink \cite{Hassink} measure a transmittance of  $0.5 - 0.6$ in energy range $10 - 30$~eV. There is no direct experimental information on absorption coefficient and in our simulations we adopt several plausible values.} $\sim 1-10$~eV, corresponding to de Broglie's wavelengths $4<\lambda<12$~\AA. At $\lambda$, the nanotube curved surface is almost identical to the tangent plane, being $R\gg \lambda$.  It is therefore reasonable to assume that electrons do not  resolve the curvature of the nanotubes, as if they were locally interacting with graphene planes.

The differential rates obtained with Eq.~\eqref{integ} are displayed in Fig.~\ref{grafici_rate} for an indicative value of the  DM mass of $M_\chi=5$~MeV.  To compute the rates we have assumed a Maxwell-Boltzmann velocity distribution $f(\bm v)$ for the DM particle velocities in the laboratory frame (see Eq.~\eqref{vdistri}).

The three curves reported in  Fig.~\ref{grafici_rate}, are relative to three different orientations $\theta_w$ of the DM wind average direction $ \bar{\bm  v}$ with respect to the carbon nanotube parallel axes (both $sp^2$ and $\pi$--electrons are considered).  

Consider a graphene plane oriented orthogonally to the DM wind. Consequent electron recoils will be oriented in the same direction.
In the case of carbon nanotubes, given the curvature of the surface the DM will collide on, the effect is slightly modified with respect to what found on  graphene sheets. Despite the fact that largest electron recoil rates are for $\theta_w\simeq 90^\circ$ (and electron recoils are in the forward region) the largest fraction of  electrons collected in the detection region corresponds to those ejected with a small angle with respect to  CNT axes. In this sense, the detector is `{\it directional}': the larger number of countings is expected  when the CNT axes are in the direction of Cygnus.

The asymmetry defined in Eq.~\eqref{asymmetry}, in absence of background, is  $A(0) \approx 0.4$.
Changing $C$ by a an order of magnitude, $A$ changes by  $\approx 10\%$. Calculations are done including both electrons from $\pi-$orbitals and from $sp^2-$hybridized orbitals.

To obtain a $5\sigma$ evidence of a non-zero asymmetry, we compute  the exposure, in units of the target mass times the data acquisition time for a fixed value of the absorption coefficient and $M_\chi=5~$MeV.  This quantity expresses the amount of target mass needed or the number of years of exposure to appreciate a statistically significant asymmetry. We find that
\be
M\cdot t (\mathrm{kg}\cdot \mathrm{day})\simeq 16 
\ee
Varying $C$ by an order of magnitude,  $M\cdot t$ varies by a approximately a factor of 2. Calculations are done including both electrons from $\pi-$orbitals and from $sp^2-$hybridized orbitals and with a $\sigma_{e\chi}\sim 10^{-37}$~cm$^2$.

We have to underscore here that in order to measure a certain degree of asymmetry $A$, we do not need to precisely measure the electron recoil direction. We only need to count the electrons reaching the detection region. 

\mysection{A compact apparatus}
We consider an array of single-wall metallic carbon nanotubes positioned in vacuum and in a  uniform electric field directed  parallel to CNT axes. CNTs are held at a fixed negative potential. Field lines will concentrate on the open ends of this CNT cathode, like on sharp edges, as described in Fig.~1 and commented in the Introduction.   Electrons  ejected by collisions with DM particles will travel in vacuum regions among (or within) CNTs and will eventually reach the region where the electric field is intense.  Once there, electrons will be further accelerated in an electric field of several ~kV/cm towards the anode where a silicon diode is located,  as in a hybrid  light sensor  (HPD  or HAPD).

The signal produced by  a collision with a single DM particle is expected to be represented by   {\it single electron count}. Therefore, the detector has to be devised to discriminate between single and multi-electron signals. This might be obtained with  HPD-type sensors,  having an intrinsically   low gain fluctuation, when coupled to a very low electronic noise amplification stage. Notice that in this configuration, given the very low rate of interaction, neither fast nor highly segmented sensors are  required.

On the other hand,  we expect photons from radioactivity to convert into the CNT target array. This would generally produce  
electrons with keV or higher energies. These events are expected to extract   
several electrons from  the CNT cathode. Therefore  the signal-to-background 
discrimination, at this level, is that between single-electron and multi-electron counts.

The  detection element can be replicated to reach the required target mass. Eventually, two  arrays of elements  can  be installed on a system that is tracking the Cygnus apparent position. 
Two CNT arrays can be installed in a back to back configuration: in one the open ends are in the direction of the Cygnus (where the DM  wind is expected to come from).
A different counting rate is then expected on the two arrays,  maximally exploiting the anysotropy of the detection apparatus.  More sophisticated schemes might require the use of magnetic and electric fields, such as the one sketched in~\cite{Hochberg:2016ntt}.

We conclude that the anisotropic response studied in this note allows to use existing technology  with the substitution of the photocathode element only, and making them blind to light. This makes our proposal easy to test experimentally and scalable to a large target mass. 
For the sake of illustration, assume a $1\times 1$ cm$^2$ substrate coupled to a single photo-diode channels. On this substrate a  number of ~$10^{12}$, 10 nm diameter CNTs  can be grown. Since the surface density of a graphene sheet is ~1/1315 gr/m$^2$,  a single-wall CNT weights about $50 \times 10^{-16}$ grams. This is equivalent to $\sim 10$~mg on a single substrate.
 In the case of HPD, O($10^4$) units per 100 g CNT are needed. In principle, the system is scalable at will, since the target mass does not need to be concentrated in a small region.
 
Single electrons counts can be triggered by environment neutrons as well. This is a well known source of 
background afflicting all direct DM search experiments and the screening techniques are the standard ones.
Thermal neutrons have scattering lengths of few fermis with electrons in graphene, but they have not enough energy 
to extract them efficiently from the material. A neutron moderation screen, as those currently used in these kind of 
experiments, has to be included when devising  the apparatus. 
We assume that working with compact units as HPDs, this kind of screening might be achieved more easily 
than with other configurations.
\newline\indent
Another source of single electron counts, which belongs to similar configurations too, is the  
electron thermo-emission. This can strongly be attenuated by cooling  the device down to cryogenic 
temperatures. However, as noted in~\cite{liang}, the {\it thermionic} electron current from an 
effective surface of 1 m$^2$ of graphene should definitely be negligible at room temperatures being proportional to\footnote{with  a coefficient $\beta =115.8~A/m^2\, K^{-3}$.} 
\be
j\approx T^3 \exp{(-\phi_{\rm wf}/kT)}
\ee
This is essentially due to the fact that the work-function $\phi_{\rm wf}$ in graphene is almost three times as large than the typical work-function of photocathodes. 
\newline\indent
As for the field emission, this has also been studied in [18] where it is found that its starts
 being significant for electric fields above 1V/nm, way larger than the ones we consider, see~\eqref{field}.  
\newline\indent

 \mysection{Conclusions}
We have shown that single wall carbon nanotube arrays might serve as directional detectors also for sub-GeV DM particles, if  an appropriate external electric field is applied and electron recoils  are studied. 
An appreciable anisotropic response, as large as $A\sim 0.4$ in~\eqref{asymmetry},  is reached with a particular orientation  orientation of the target with respect to the DM wind. Since the proposed detection scheme does not require any precise determination of the electron ejection angle and recoil energy, the carbon nanotube array target could be integrated and tested in a compact Hybrid Photodiode system | a technology already available | made blind to light.  High target masses can be arranged within limited volumes with respect to configurations proposing to use graphene planes. 

The results presented are obtained starting from the conclusions reached by Hochberg {\it et al.}~\cite{Hochberg:2016ntt} on DM scattering on graphene planes and adapted to  the wrapped configuration of single wall carbon nanotubes. The fact that carbon nanotubes, and interstices among them in the array, almost behave as empty channels is still an essential feature to obtain  the results of the calculations described here. The mean free paths attainable in these configurations are definitely higher if compared to dense targets as graphite or any crystal.  We also observe that, in the detection scheme  proposed, differently from~\cite{Capparelli:2014lua}, small irregularities in the geometry of nanotubes are inessential. 

\begin{figure}[hbt!]
\centering
\includegraphics[width=7truecm]{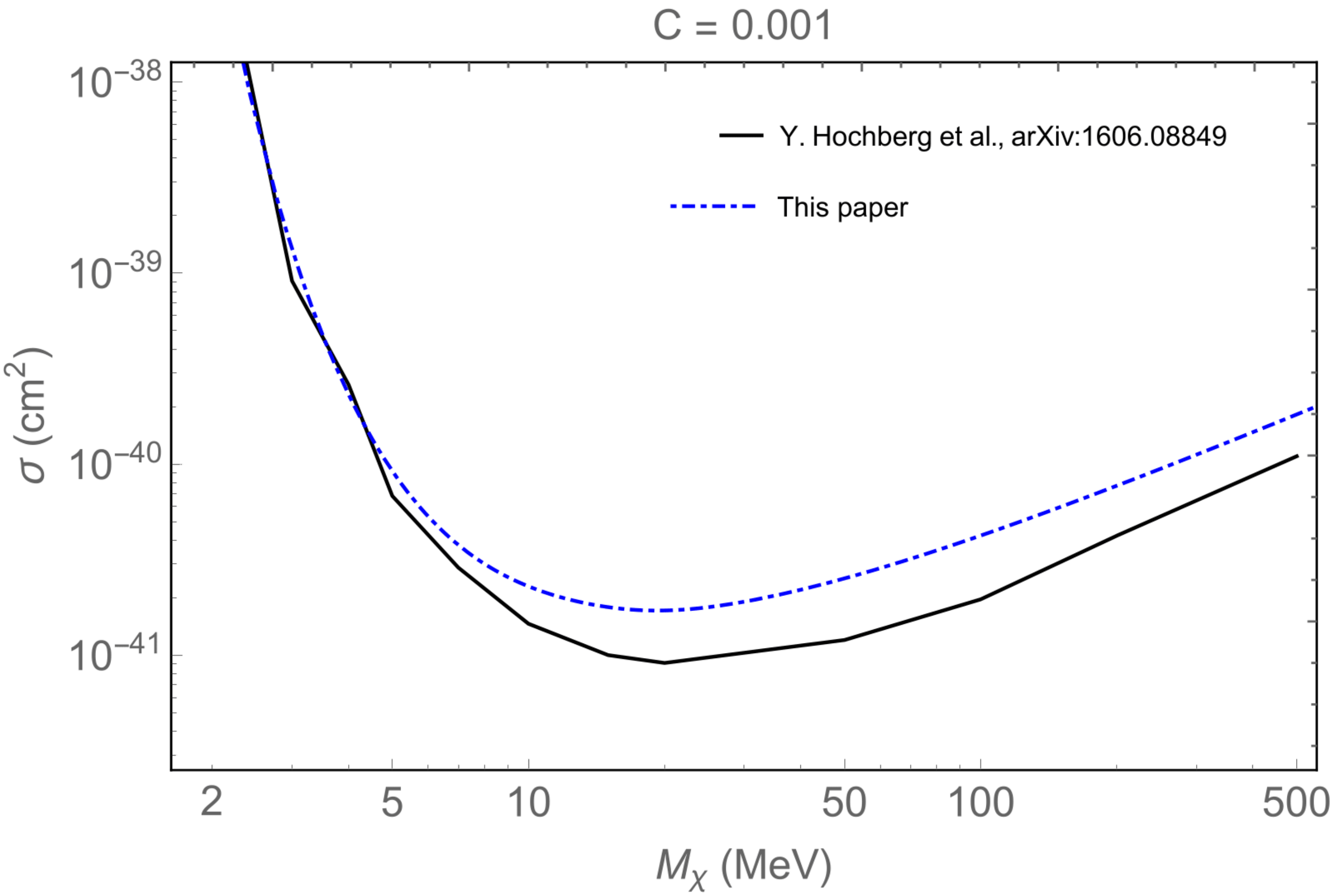}
\caption{\small We compare our results  with those obtained by Hochberg {\it et al.}~\cite{Hochberg:2016ntt}.
Calculations are done including both electrons from $\pi-$orbitals and from $sp^2-$hybridized orbitals. The exposure of 1~kg$\times$year is used.  }\label{Exclusionlimits}
\end{figure}

For comparison with previous work, we present the exclusion plot, see Fig.~\ref{Exclusionlimits}, which can be obtained with the detection configuration here proposed. We perform a full calculation including $\pi$ and $sp^2-$ electrons. 
The latter figure summarizes the potentialities of the scheme proposed. They result to be very much comparable to what found in~\cite{Hochberg:2016ntt}, although with rather different apparatus and   practical realization.  To conclude, we notice that the device here described might be used alternatively as a detector of heavier DM particles. Just by changing the direction of the electric field, one could count positive carbon ions recoiled out of and channeled by the carbon nanotubes (or within the interstices among them), as in the original proposal~\cite{Capparelli:2014lua}~\cite{Cavoto:2016lqo}.

\mysection{Acknowledgements} We are very grateful to Yonit Hochberg for several comments and suggestions on the manuscript and to Chris Tully for informative discussions. 
We also thank  Maria Grazia Betti,   Carlo  Mariani and Francesco Mauri for several useful hints on the physics of CNTs. We thank an anonymous referee for extremely useful comments and suggestions.  G.C. acknowledges partial support from ERC Ideas Consolidator Grant CRYSBEAM G.A. n.615089.

\mysection{Appendix: DM-electron scattering}
In this Appendix we report the essential formulae we have used to obtain the results in the text. We have adapted the expressions in~\cite{Hochberg:2016ntt} to the configuration with CNTs. 

The  $M_\chi$ DM mass needed to eject electrons from graphene  is about $3 \MeV$ at the galactic  escape velocity. In  the $\chi e^-$ scattering process, part of the momentum is exchanged with the crystal lattice --- this is especially true for exchanged momenta smaller than the inverse of the average spacing between atoms  $ a = 0.14 \, \mathrm{nm} = 0.721\, \mathrm{keV}^{-1}$. 
 
 The energy released by the DM particle in the scattering process  is
 \be
\Delta E = E_i(\bm \ell) + \phi_{\rm wf}+ k^{\prime 2} /2m_e
\ee
where  $-E_i(\bm \ell)$ is the energy of the electron in the valence band (depending on the lattice momentum $\bm \ell$), $-\phi_{\rm wf}$ is the work function of graphene and $k^\prime$ is  the momentum of the ejected electron.   Let $q$ be the DM four-momentum difference  before and after the scattering. One finds that 
\be
\Delta E = \bm v\cdot \bm q - \frac{q^2}{2M_\chi}
\ee
Therefore the cross section of  the $\chi e^-$ scattering is
\bea
d\sigma_{ \ell} &=& \frac{1}{F} |M_{e\chi}(q)|^2 \frac{d^3 p^\prime}{(2\pi)^3 2\varepsilon^\prime}\frac{d^3 k^\prime}{(2\pi)^3 2E^\prime} |\widetilde{\psi}(\bm q-\bm k^\prime,\bm \ell)|^2\no\\ &\times& (2\pi) \delta \left( E_i(\bm \ell)+ \phi_{\rm wf} + \frac{k^{\prime 2}}{2m_e} - \bm v \cdot \bm q + \frac{{q}^2}{2M_\chi} \right)
\eea
where $F$ is the flux $F = 4  \varepsilon E |v| = 4 M_\chi m_e |v|$ and $M_{e \chi}(q)$ is the amplitude of the transition process.  Electrons in $\pi-$ orbitals have rather soft kinetic energies which allow $E\simeq m_e$. Here $p^\prime$ and $\varepsilon^\prime$ are related to the DM after the collision with the electron.  The factor $|\widetilde\psi(\bm q-\bm k^\prime,\bm \ell)|^2$ measures the probability for the recoiled electron to have 3-momentum $\bm k^\prime$ for $\bm q,\bm \ell$ fixed~\cite{Hochberg:2016ntt} (and has dimensions of eV$^{-3}$). 
We also consider electrons from $sp^2-$hybridized orbitals.  The explicit forms of $E_i(\bm \ell)$ and  $\widetilde \psi(\bm q-\bm k^\prime,\bm \ell)$ are different for the two cases~\cite{Hochberg:2016ntt}; see also~\cite{book}. The calculations are done separately for the four orbitals (three in the $sp^2-$hybridized configuration). Electrons in  $\pi-$orbitals are more weakly bound and more sensible to light DM particles.  At higher recoil energies $\sigma$-electrons dominate.

The definition is used
\be
 \frac{|M_{e \chi} (\alpha^2 m_e^2)|^2}{16\pi m^2_e M^2_\chi} \equiv \frac{\sigma_{e\chi}}{\mu^2_{e\chi}}
\label{sigele}
 \ee
where $\sigma_{e\chi}$  is the cross section of the non-relativistic $\chi e^-$ elastic scattering and $\mu_{e\chi}$ is the reduced mass of the electron-DM system.  In most calculations we use $\sigma_{e\chi}\simeq 10^{-37}$~cm$^2$ as a benchmark value for the cross section. It is found
\begin{eqnarray}
\sigma_\ell &=& \frac{\sigma_{e\chi}}{\mu^2_{e \chi}} \frac{1}{2(2\pi)^4 |v|}\int d^3 p^\prime \, d^3 q \, |\widetilde{\psi}(\bm q-\bm k^\prime,\bm \ell)|^2 \no\\
&\times&\delta \left( E_i(\bm \ell)+ \phi_{\rm wf} + \frac{k^{\prime 2}}{2m_e} - \bm v \cdot \bm q + \frac{{q}^2}{2M_\chi} \right)
\label{xsect}
\end{eqnarray}

The Dirac delta function defines a minimum speed for  $\chi$ to  eject  the electron
\be
v_{\rm min} = \frac{\Delta E}{|\bm q|} - \frac{|\bm q|}{2M_\chi}
\label{vminima}
\ee
If the minimum speed were higher than the Milky Way escape velocity ($v_{\rm min} > v_{\rm esc} +v_0 = 550+220~\mathrm{Km/sec}$) the process would simply be forbidden. 
If we assume that $|\bm q|\ll M_\chi$, then
\be
\frac{\Delta E}{|\bm q|} \simeq v_{\rm min}< v_{\rm esc}+v_0
\label{dis1}
\ee
which in turn means 
\be
|\bm q|\gtrsim\frac{4.3\eV}{v_{\rm esc}+v_0}\simeq 1.7\keV
\label{dis2}
\ee

Observe also that $|\bm q|< a^{-1}=8.7 \keV$ so that 
\be
1.7\keV\lesssim |\bm q|\lesssim 8.7\keV
\ee

The total rate, per unit of time and  detector mass is 
\begin{equation}
\label{Rate_finale}
R = N_C \frac{\rho_\chi}{M_\chi} \, A_{cu} \int_{\bm \ell \in B_1} \frac{d^2\ell }{(2\pi)^2}\, d^3v\,  \,f(\bm v) \,v\,\sigma_\ell
\end{equation}
with $A_{cu} = \frac{3\sqrt{3}}{2}a^2$ unit cell area of the graphene and $B_1$ the first Brillouin zone of the reciprocal lattice\footnote{
Observe that $\int_{\bm \ell \in B_1} \frac{d^2 l}{(2 \pi)^2} = \frac{1}{A_{cu}}$. 
}. $N_C$ is the number of carbon atoms per kg, $N_C=5\times10^{25}$~kg$^{-1}$. $\rho_\chi/M_\chi$ is the DM number density, with $\rho_\chi\simeq0.4$~GeV/cm$^3$ being the local density. Finally $f(\bm v)$ is the velocity distribution to be defined below. 

In the specific case of single wall carbon nanotubes, which is of interest in this paper, periodic boundary conditions are imposed on the argument  $\bm\ell$ of $|\widetilde\psi(\bm q-\bm k^\prime,\bm \ell)|^2$. 
If $x$ is the coordinate on the boundary of the nanotube, and $r$ is its radius, at fixed altitude $z$, the condition
\bea
\exp(i(x+ 2\pi r)\ell_x)=\exp(ix\ell_x)
\eea
leaves $\ell_y$ continuous in the integral~\eqref{Rate_finale} whereas a discrete sum on $\ell_x$ has to be taken
\be
\int d^2\ell\to \sum_{n}\int d\ell_y
\ee
where $\ell_x=n/r$, in the first Brillouin zone.  In the following this substitution is understood. 

Replacing  the cross section formula for  $\sigma_\ell$ in~\eqref{Rate_finale} it is found
\begin{eqnarray}
R &=& N \int d^2 \ell\int d^3 k^\prime\, d^3q \; |\widetilde{\psi}(\bm q-\bm k^\prime,\bm \ell)|^2\nonumber \\
 &\times&\int_{v_{\rm min}(\bm \ell, k^\prime, q )}^{v_{\rm max}}d^3v \, f(\bm v)\, \delta(v_{\rm min}|\bm q| - \bm v \cdot \bm q)
\label{ratetc}
\end{eqnarray}
where 
\be
N=\frac{1}{2(2\pi)^6}\frac{N_C \rho_\chi A_{cu}}{M_\chi } \frac{\sigma_{e\chi}}{\mu^2_{e \chi}}
\ee
and
\be
v_{\rm min}(\bm \ell, k^\prime, q) = \frac{ E_i(\bm \ell)+\phi_{\rm wf}+ \frac{k'^2}{2m_e}}{|\bm q|}-\frac{|\bm q|}{2M_{\chi}}
\ee
In Eq.~\eqref{ratetc} $v_{\rm max}$ is computed solving the inequality
\be
v(v-2v_0 \cos\theta)\leq (v_{\rm esc}^2-v_0^2)
\ee
and
\be
d^3v=v^2\, dv\, d\cos\theta\,d\phi
\ee

We can turn to the differential rate in the recoil energy. Differentiating the recoil energy of the electron $E_r = \frac{k'^2}{2M_\chi}$ one has $\frac{d R}{d \ln E_r} =E_r \frac{d R}{d E_r} = \frac{k^\prime}{2} \frac{d R}{d k^\prime}$ which allows to write the differential cross section distribution
\begin{eqnarray}
&&\!\!\!\!\!\!\!\!\! \mkern-18mu \frac{d R}{d \ln E_r} = N \int k^{\prime 3} d\Omega_{k'}  \int  d^2\ell \int_{q_{\rm min}}^{q_{\rm max}}\!\!\!\!\!\!\!\!  d\phi_q\, dq |\bm q| \, |\widetilde{\psi}(\bm q-\bm k^\prime,\bm \ell)|^2  
\notag\\
&\times&\!\!\! \mkern-6mu\int_{v_{\rm min}(\bm \ell,k^\prime, q)}^{v_{\rm max}} \!\!\!\!\!\!\!\!\!\!\!\!\!\!  d \Omega_v \, dv\, v \, f(\bm v) \int d\cos\theta_q\, \delta \left( \cos\theta_q - \frac{v_{\rm min}}{v} \right)
\label{integ}
\end{eqnarray}
where $\theta_q$ is the angle between $\bm q$ and $\bm v$ (and $\phi_q$ the azimuthal angle around $\bm v$)
\be
\delta(v_{\rm min}|\bm q| - \bm v \cdot \bm q) = \frac{1}{|\bm q|v}\delta\left(\cos\theta_q-\frac{v_{\rm min}}{v}\right)
\ee
The Maxwell-Boltzmann  distribution of velocities $f(\bm v)$ in the Galaxy is 
\bea
f(\bm v) = \alpha\, e^{-\frac{(\bm v-\bm v_0)^2}{v_0^2}} \theta(v_{\rm esc}-|\bm v-\bm v_0|)
\label{vdistri}
\eea
where $v_0\simeq 220$~Km/s with $\bm v_0$  directed towards the Cygnus constellation.  The carbon nanotubes axes may 
have different orientations with respect to the DM wind $\bm v_0$ vector. Once $\bm v_0$ is fixed, a weighetd sum with $f(\bm v)$ of the directions (and lengths) of $\bm v$ around $\bm v_0$ is taken.  For each term in the sum, the direction of $\bm q$ is fixed, as in~\eqref{integ}. Changing the orientation of the carbon nanotube axes  with respect to the Cygnus ($\bm v_0$ w.r.t. the parallel axes), the distribution   $dR/d\ln E_r$  changes as shown in Fig.~\ref{grafici_rate}. 

On the basis of~\eqref{dis1} and using as a minimum value for $\Delta E$ the work function $|\phi_{\rm wf}|\sim 4$~eV and as typical $|\bm q|$ values $2<|\bm q|<8$~keV, we see that $v_{\rm min}\approx v_0\sim 10^{-3}$. 

\begin{figure}
\includegraphics[width=8.5truecm]{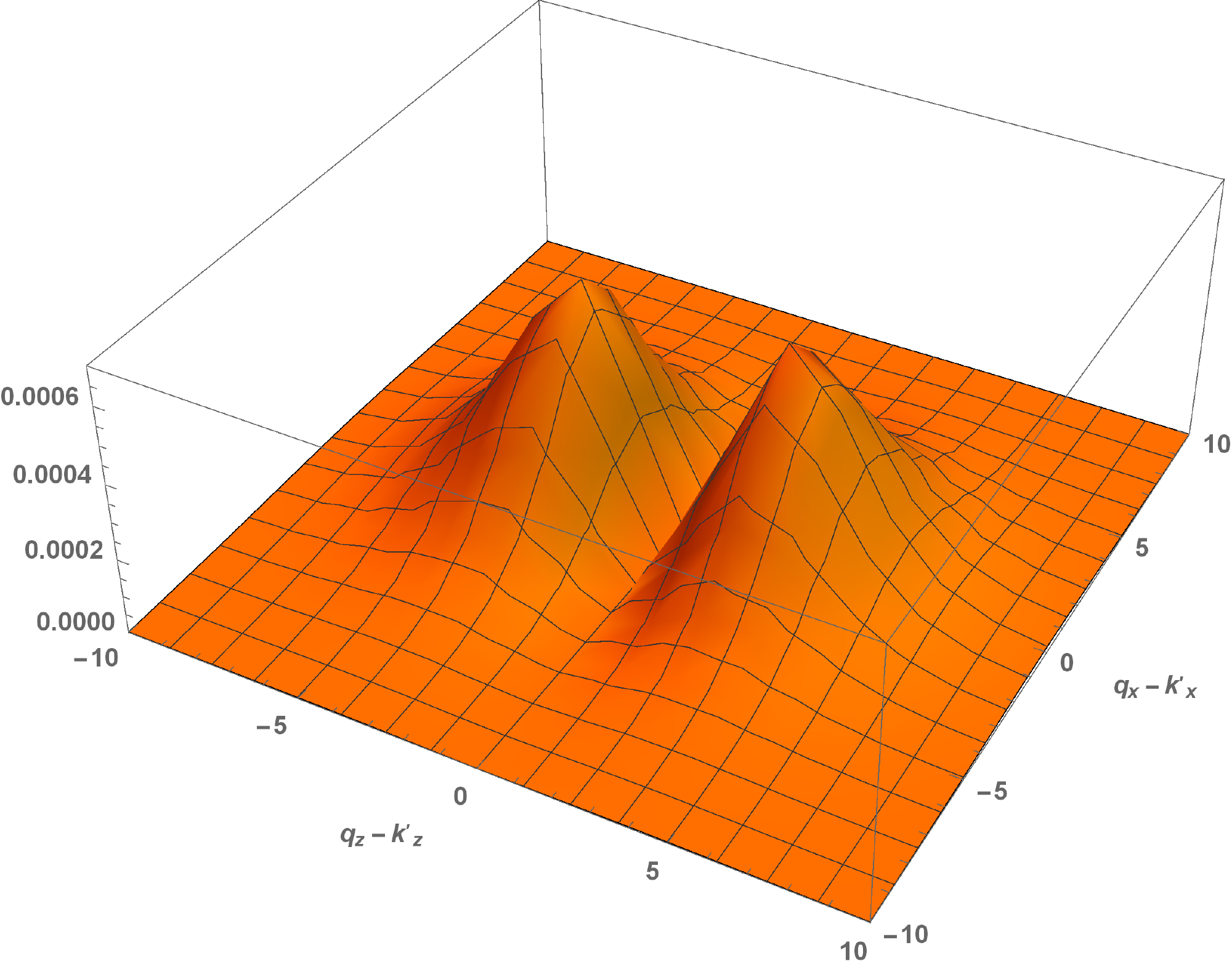}
\caption{\small The case of a $\pi-$orbital. $|\widetilde{\psi}(\bm q-\bm k^\prime,\bm \ell=0)|^2$ (center of Brillouin zone). Figure is plotted as a function of $\bm q-\bm k^\prime$ in keV. 
$k_x^\prime$ is in the lattice plane whereas $k_z^\prime$ is orthogonal to the lattice. As can be seen from the figure, it is unlikely to have a $q-k_z^\prime\approx 0$, $|\widetilde\psi|^2$ measuring the probability for the ejected electron to have $\bm k^\prime$ momentum, once $\bm q$ and $\bm \ell$ are fixed. 
}\label{figura_psi}
\end{figure}

If the DM is orthogonal to the graphene plane, then the largest component of $\bm q$ will be in same direction.  For the sake of illustration, set  $\ell=0$,  and  make reference to Fig.~\ref{figura_psi}.  Then we see that the electron is most likely recoiled with $k^\prime_z$ values not too different from $q_z$ and with  small $k^\prime_{x,y}$ values --- the $z$ direction is the one orthogonal to the graphene plane.  This in turn means that the ejected electrons tend to follow the same direction of the incoming DM particle.

\end{document}